# ROSAT observations of the galaxy group AWM7

D.M. Neumann & H. Böhringer

Max-Planck-Institut für extraterrestrische Physik, D-85740 Garching bei München, Federal Republic of Germany

**Abstract.** We present results of ROSAT/PSPC and HRI observations of the AWM7 group of galaxies, which is a poor galaxy cluster and forms part of the Perseus-Pisces filament. The X-ray emission originates from intracluster gas at temperatures of 1.7 to 4.5 keV. The cluster obviously is elliptical with a position angle perpendicular to the position angle of the dominant elliptical galaxy NGC1129, which is offset from the cluster X-ray centre by 30 kpc. The analysis of the PSPC imaging and spectral data yield a gravitational mass of $2-5 \cdot 10^{14} M_\odot$ within a radius of 1.2 Mpc and a cooling flow with a mass deposition rate of up to 60–66 $M_\odot\,y^{-1}$.

## 1. Introduction

AWM7 is a group of galaxies found by Albert, White and Morgan (1977) during a search for galaxy groups around cD galaxies in the National Geographic Society–Palomar Observatory Sky Survey. It is located in the vicinity of the Perseus cluster and actually forms part of the Perseus-Pisces galaxy filament (Giovanelli et al. 1986; Focardi et al. 1984). AWM7 would be classified as a richness class 0 galaxy cluster (Kriss et al. 1983) but was not included in the Abell catalogue (Abell 1958) due to its close vicinity to the galactic plane. So far there are 46 galaxies known to belong to the cluster within a radius of $1.5°$ (Dell'Antonio et al. 1994). For 33 galaxies redshifts are listed in Beers et al. (1984, hereafter BGHLD). The redshift of the central cD galaxy of AWM7 is 0.0176 (BGHLD; Stauffer & Spinrad 1980).

While the typical mass-to-light ratio of rich galaxy clusters, deduced from a virial analysis or from X-ray data (e.g. Kent & Sargent 1983; Hughes 1989; Henry et al. 1993), is 100 - 400 $M_\odot/L_\odot$, BGHLD have found a mass-to-light ratio in the range of 850 - 1400 $M_\odot/L_\odot$ from a mass analysis of AWM7.

X-ray observations provide an especially good tool for the mass determination and allow us to reinvestigate this exceptionally high mass-to-light ratio. With ROSAT (Trümper 1983, 1992) it is now possible to study the morphology of the intracluster medium (hereafter ICM) of this group of galaxies in more detail. First, the PSPC (Position-Sensitive-Proportional-Counter) facilitates the opportunity to study simultaneously the spectral and spatial properties of X-ray sources to an accuracy as yet unknown in X-ray astronomy. Secondly, the HRI (High-Resolution-Imager) offers the highest spatial resolution reached so far in X-ray telescopes.

Earlier X-ray observations (Canizares et al. 1983; Kriss et al. 1983; Edge et al. 1992) showed evidence for a cooling flow in the central region of AWM7. With the new data of ROSAT, the existence of a cooling flow is confirmed. Edge et al. (1992) determined a rate of accreting material up to 56 $M_\odot$ within a radius of 65 kpc.

In the following section an account of the observations is given. Section 3 describes the morphological analysis with particular attention to the surface brightness profile and the ellipticity of the cluster. We further present evidence for substructure. In section 4 results of the spectral analysis are given, yielding a radial temperature profile for AWM7. In our analysis we found an excess of absorbing hydrogen in the cooling flow region.

In section 5 we present the results of a new, model-independent method of mass determination. The results of a cooling flow analysis are also presented. In section 6 we discuss the results and list our conclusions. $H_0 = 50$ km s$^{-1}$Mpc$^{-1}$ is used throughout this paper.

AWM7 was observed with the ROSAT/PSPC for 13,335 seconds and for 14,864 seconds with the HRI. Figures 1 and 2 show the resulting images of the two exposures. The PSPC observation yielded approximately 170,000 photons, while the HRI exposure provides about 100,000 photons in total. In both observations the centre of overall symmetry coincides with the centre of the detector. The spatial resolution of the PSPC is $\simeq 30''$ (0.4-2.4 keV FWHM), while that of the X-ray telescope - HRI combination is $\simeq 5''$. The energy resolution of the PSPC is about 40% at 1 keV (Briel et al. 1988).

The observations were made in January 1992 and July 1992, respectively. The PSPC image (Fig. 1) shows only the counts in the energy range 0.4-2.4 keV (the full range of the ROSAT telescope is 0.1-2.4 keV). As the ICM has normally very high temperatures ($10^7 - 10^8$ K), the energy is mostly emitted in a range between 1-10 keV. Therefore, the soft ROSAT band ($\leq 0.4\text{keV}$) is strongly contaminated by background originating mostly from the galactic plane. For that reason we used only the hard photons for the PSPC image. The HRI image is made up of photons from the full energy range of the instrument, as this detector has no energy resolution. Both images are filtered with a variable Gaussian filter. The filter is broader for the regions of lower surface brightness and gets systematically narrower for brighter regions. Therefore, it is possible to get information on structures on different scales, from the overall ICM distribution to individual point sources. The filter is always broader than the spatial resolution of the two detectors.

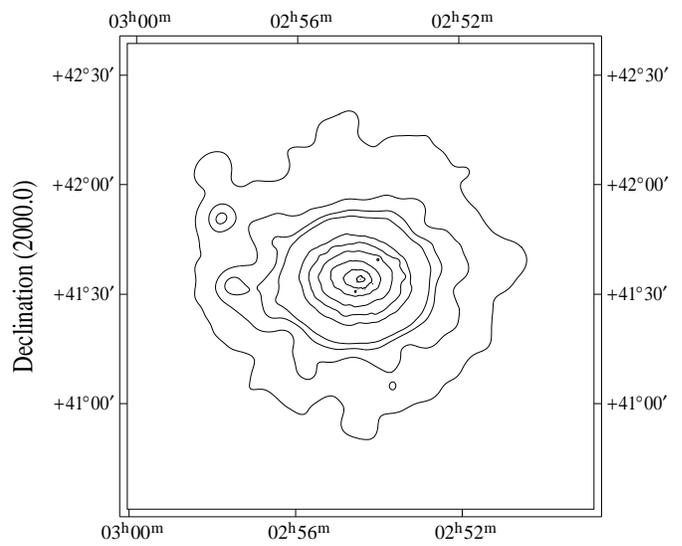

**Fig. 1.** ROSAT/PSPC image of AWM7 in the hard ROSAT band (0.4-2.4keV). The image is filtered with a variable Gaussian filter which gives the possibility to see structures on different scales. The largest FWHM of the filter is 23'. The plotted photon count levels are: 0.2,0.4,0.6,0.8,2.,3.,5.10., 20.,50.,150. The background intensity is 0.12-0.15.

## 3. Morphological data analysis

An interesting aspect in morphological studies of the ICM is substructure, as the gas traces the shape of gravitational potentials, if the cluster is close to hydrostatic equilibrium and isothermal.

Figure 1 and 2 show interesting morphological features of this galaxy cluster. At a radius larger than $\sim 80$kpc, the isophotes are very well characterized by ellipses with a ratio of minor to major axis of about 0.8. The major axis is oriented in east-west direction. The inner part of the cluster (Fig. 2) shows a shift of the X-ray maximum to the west. This maximum coincides exactly with the central dominant cD galaxy NGC1129. The region where the isophotes are shifted to the west cover approximately the area of the cooling flow as will be discussed below.

We derived X-ray surface brightness profiles of AWM7 from both the PSPC and HRI data. The azimuthal average of the surface brightness derived from the PSPC data with the central position on NGC1129 is shown in Fig. 3. Deviations in symmetry, as measured by the ellipticity of the cluster are neglected in the averaging process.

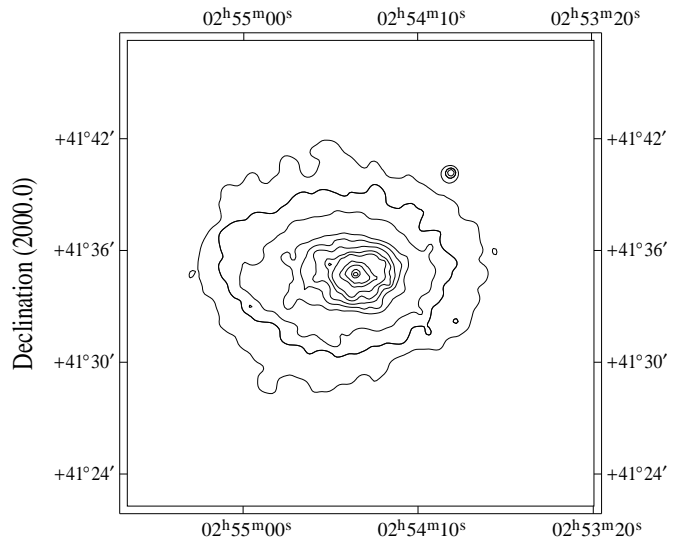

**Fig. 2.** ROSAT/HRI image of AWM7 in the broad band. The resolution is higher than in the PSPC image. The image is smoothed with a variable Gaussian filter. The largest FWHM we used is 5.3', the photon count levels are: 2.,2.5,3.,3.5,4.,4.5,5.,5.5,6.,8.,10.,15.,20. The background intensity is 1.0-1.1.

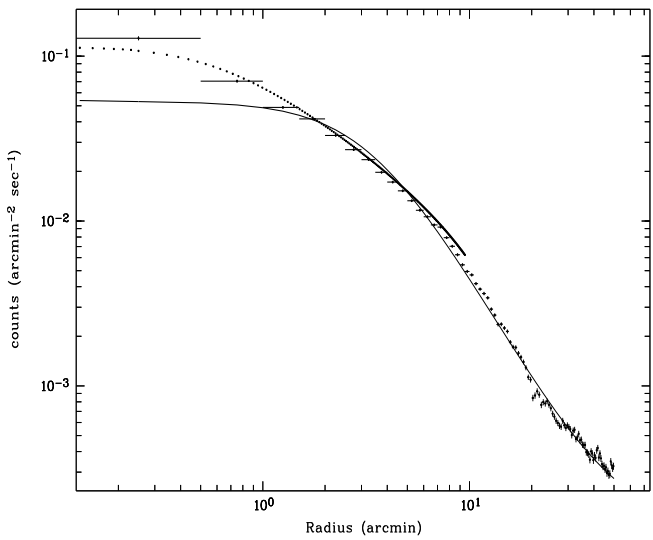

**Fig. 3.** Surface brightness profile of AWM7 from the PSPC data. The full line shows the overall fit of the isothermal $\beta$-model. The dotted line is the result for the central fit, derived from the HRI data.

The surface brightness profile shown in Fig. 3 was fitted by an isothermal "$\beta$-model" (Cavaliere & Fusco-Femiano 1976; 1981; Sarazin & Bahcall 1977; Gorenstein et al. 1978; Jones & Forman 1984). The surface brightness has the functional form

$$S(r) = S_0 \left(1 + \frac{r^2}{a^2}\right)^{-3\beta + \frac{1}{2}}. \qquad (1)$$

$S_0$ is the central brightness, $a$ is the so-called core radius (the radius at which the brightness has reached approximately half the central value), $\beta$ is a slope parameter, originating from the ratio between kinetic energy of the galaxies and the thermal energy of the gas. In Fig. 3 two $\beta$-model fits to the data are shown. An overall fit to the PSPC data provides a good representation of the surface brightness profile outside a radius of 2' (~60kpc) but underestimates the intensity in the very centre of the cluster. A fit of a model constraint to the central part of the HRI data gives a good representation of the inner region. The latter analytical approximation was used for the cooling flow analysis described below. For the following mass analysis we were mainly interested in the distribution of the gas mass and gravitational mass on larger scales and therefore used the first $\beta$-model fit, which is a valid description of the data outside 60kpc. The error in the gas density introduced by the overall fit is about 5% and therefore negligible to the errors in temperature, which are shown in Table 1. The surface brightness (Eq.(1)) can be analytically deprojected – under the assumption that the cluster is spherically symmetric. Since the emissivity of the ther-2-10keV shows only small variations, we can neglect temperature variations with an insignificant loss of accuracy and derive directly the gas number density distribution $n_g$, given by

$$n_g = n_{g0} \cdot \left(1 + \left(\frac{r}{a}\right)^2\right)^{-\frac{3}{2}\beta}. \qquad (2)$$

In our analysis we find $\beta=0.53 \pm 0.01$, $a=102\pm5$kpc, and $n_{g0}=0.012$–$0.013$cm$^{-3}$. The central fit leads to $a=3$–$8$kpc, $\beta=0.25\pm0.01$, and $n_{g0}=0.06$–$0.07\pm$cm$^{-3}$.

An optical CCD image of the central dominant galaxy NGC1129 is shown in Fig. 4. This was taken at the Calar Alto 1.23m telescope, through a red filter with an exposure time of 300 s. Peletier et al. (1990) have found that NGC1129 has a strong twisting in the Position angle of the isophote levels, excluding the chain of four galaxies which seems to be related to the cD galaxy. The overall ellipticity, however, seems to have a major axis in north-south direction which is perpendicular to the overall ellipticity of AWM7.

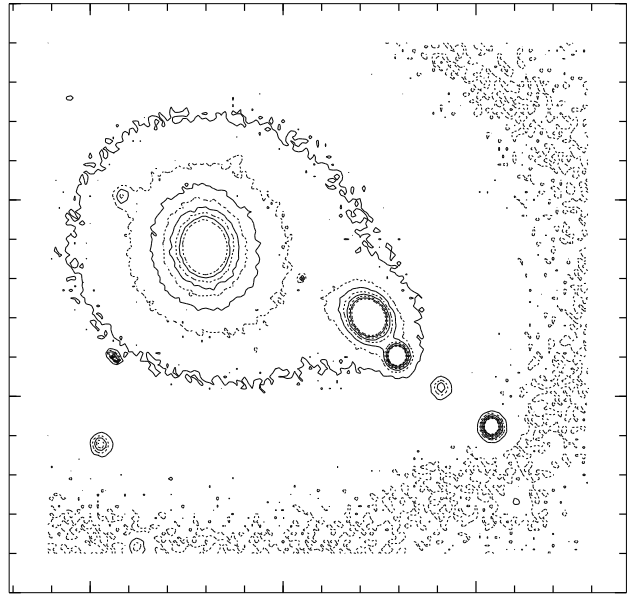

**Fig. 4.** NGC1129 with a chain of four satellite galaxies. The exposure was taken at the 1.23m telescope at Calar Alto. The length scale of the image is ~80".

## 4. Spectral analysis

The energy resolution of the ROSAT/PSPC provides the opportunity of measuring temperatures of the ICM. We determined temperatures in a series of concentric rings around NGC1129. Table 1 shows the projected temperatures in the rings. Our criteria for the choice of the annuli course, to get a detailed profile; secondly, the statistics should be reasonably good, so that the temperatures are well determined (in the innermost ring we counted 3,000 photons, in the next about 5,000, in the following ones 10,000, in the second outermost ring about 20,000, and in the outermost ring about 70,000 photons). The number of photons is raising with increasing radius because the outer regions are more affected by background due to the decreasing surface brightness (see Fig. 3). For the temperature determination we used a Raymond & Smith code (Raymond & Smith 1977).

The limited spectral resolution of the ROSAT/PSPC and the fact that the spectra of thermal plasmas change very little as a function of temperature in the spectral range of the telescope (0.1-2.4keV), make the temperature determination rather difficult. In addition to the temperature, metallicity, hydrogen column density, and the normalization are all parameters required for the spectral analysis. Therefore, it helps if some fitting parameters are already known from other observations. In the data reduction we fixed the metallicity for all annuli to z=0.47, which is the result obtained by Ginga (Hatsukade 1989). Even though Ginga does not provide the spatial resolution to bin the data in different concentric rings, now ASCA results show evidence for a metallicity gradient in the ICM (Ikebe et al. 1994; Ohashi et al. 1994), but is too small to affect significantly the spectral analysis.

We binned the spectra obtained by the ROSAT/PSPC to a signal-to-noise ratio of three and applied a $\chi^2$ fit with the temperature, the hydrogen column density, and the normalization as fit parameters. The most distant annulus overlaps with the next inner annulus to obtain a better quality of the spectrum.

As background we used a ring between a radius of 42'-50', where the emission of the ICM is almost negligible. For the spectral analysis only the data in the energy band 0.2-2.4 keV were used and for the largest annulus the fitting was restricted to the 0.41-2.4 keV energy band, due to the much poorer statistics. The low energy channels had to be discarded because the standard vignetting correction overcorrects at these channels and introduces an error in the spectra of the outer annuli. This limitation affects the accuracy of the determination of the hydrogen column density but does exclude systematic errors from the spectral analysis (e.g. prevents a too high galactic absorption value).

For our analysis we used a one-temperature model for every annulus. We did not correct for rings outside the regarded annulus, which overlap in the line of sight because the main part of the emission comes from the shell we look at. Multi temperature fits lower the statistics, which leads to a decrease of accuracy shown in a second analysis described below. The temperatures we derived are in good agreement with the Ginga results of David et al. (1993), who derived an overall temperature of 3.9–4.2 keV (1-$\sigma$ confidence levels). The decrease of the temperature in the centre of AWM7 confirms the existence of a cooling flow. In a second analysis we made multiple temperature fits to correct for the contribution of outer shells in the line of sight. The results are also given in Table 1. To account for the contributions of the outer shells to the spectra in rings we determined the temperatures from the outermost ring inwards. For each ring a multicomponent fit was performed with the previously determined temperatures and the relative contributions of the shells to the emission measure (as computed from a numerical deprojection analysis) taken as fixed parameters. The results are somewhat different to what one would expect. Instead of a steeper decrease of temperature in the central region the decrease becomes shallower. This increase of temperature in the inner ring compared to the result of the projected analysis is due to the much poorer statistics. Poorer statistics systematically increases the fit temperature, which is obviously the case here. But, however, within the error bars the two results are still consistent.

**Table 1.** Results of the spectral analysis. Line-of-sight temperatures ($T^p$) and the hydrogen column density are given with their 1-$\sigma$ errors. Also deprojected temperatures ($T^d$) from a multiple temperature fit are given.

| radius (kpc) | | | $T^p$ (keV) | $T^d$ (keV) | nH ($10^{20}$cm$^{-2}$) |
|---|---|---|---|---|---|
| 0 | – | 25 | $1.8^{+0.3}_{-0.1}$ | $2.0^{+0.2}_{-0.2}$ | $1.5^{+0.2}_{-0.2}$ |
| 25 | – | 50 | $2.3^{+0.3}_{-0.2}$ | $2.3^{+0.3}_{-0.2}$ | $1.0^{+0.2}_{-0.1}$ |
| 50 | – | 76 | $2.4^{+0.3}_{-0.2}$ | $3.4^{+0.6}_{-0.5}$ | $1.1^{+0.1}_{-0.1}$ |
| 76 | – | 126 | $3.0^{+0.4}_{-0.2}$ | $3.0^{+0.4}_{-0.4}$ | $1.2^{+0.1}_{-0.1}$ |
| 126 | – | 176 | $3.3^{+0.4}_{-0.3}$ | $3.0^{+0.5}_{-0.4}$ | $1.1^{+0.1}_{-0.1}$ |
| 176 | – | 227 | $3.2^{+0.4}_{-0.3}$ | $3.9^{+1.1}_{-0.7}$ | $1.0^{+0.1}_{-0.1}$ |
| 227 | – | 302 | $2.9^{+0.4}_{-0.2}$ | $3.0^{+0.8}_{-0.5}$ | $1.1^{+0.1}_{-0.1}$ |
| 302 | – | 378 | $3.2^{+0.4}_{-0.3}$ | $3.3^{+1.3}_{-0.8}$ | $1.1^{+0.2}_{-0.1}$ |
| 378 | – | 756 | $3.4^{+0.5}_{-0.3}$ | | $1.3^{+0.1}_{-0.2}$ |
| 378 | – | 1260 | $3.6^{+1.2}_{-0.6}$ | | $1.2^{+0.2}_{-0.2}$ |

The value for the absorbing hydrogen column density agree very well with the 21cm line data (by Dickey & Lockman 1990) with a value of $9.2 \times 10^{20}$ cm$^{-2}$ for this location close to the galactic plane. An exception is the innermost bin where a significant ($\sim$2-3$\sigma$) excess absorption is found. The good agreement between the present hydrogen column densities and the 21cm data are a confirmation of the good approximation of the one-temperature model.

## 5. Mass determination and cooling flow analysis

### 5.1. Mass determination

Under the assumption of hydrostatic equilibrium and spherical symmetry of the ICM, it is possible to calculate

data (see also Nulsen & Böhringer 1995). The necessary equations are the Euler equation

$$\frac{dP}{dr} = -n \cdot \mu m_p \frac{d\Phi}{dr}, \qquad (3)$$

and the ideal gas equation

$$P = nkT. \qquad (4)$$

$P$ is the pressure, $\Phi$ is the gravitational potential, $n$ is the number density of the gas, $T$ the temperature, and $k$ is the Boltzmann constant. Inserting Eq.(4) into Eq.(3) gives the integrated mass profile

$$M(r) = -\frac{r^2 kT}{G \mu m_p} \left( \frac{1}{\rho_g} \frac{d\rho_g}{dr} + \frac{1}{T} \frac{dT}{dr} \right), \qquad (5)$$

where, $\rho_g$ is the gas density. Including Eq.(2) gives

$$M(r) = -\frac{kr^2}{G \mu m_p} \left( -\frac{3\beta r T}{r^2 + a^2} + \frac{dT}{dr} \right). \qquad (6)$$

As the gas density profile is determined very accurately in the outer regions (see Fig. 3 - we just use the overall $\beta$-model fit from the PSPC), the main error in the mass determination is caused by the large uncertainty of the temperature; the large error bars in the temperature also allow a large range of temperature gradients.

To get a direct transformation of the uncertainties in temperatures into errors of the total mass, we developed a Monte-Carlo method, which explores the mass results for the whole range of allowed temperature profiles. The allowed temperature regime, or temperature "channel", is defined by the X-ray data (Fig. 5). A series of temperature profiles are now selected as "trajectories" through the temperature "channel" by means of a Monte-Carlo method. To model the large-scale mass distribution of AWM7, we neglect the central data within a radius of 100 kpc. Therefore the temperature drop in the three inner annuli was neglected, as can be seen in Fig. 5. This was necessary as the overall gas model, that we use (we apply the "small" fit on the cooling flow analysis), does not fit in that region and does therefore not produce the correct physical result. The applied model does not have a density gradient steep enough to compensate the rising temperature. Therefore, instead of a realistic mass determination in the cooling flow region one gets a decreasing or even a negative total mass. To overcome this problem we extrapolated the temperature of the forth annulus to the centre.

In the analysis mass profiles for 1000 successful trajectories were calculated. Then the masses of the trajectories were sorted according to their mass at each radius. We used the 10% and the 90% limits of the masses at each step as errors. We neglected the upper and lower 10% values due to the fact that they resulted from trajectories with especially high gradients. The results are shown together with the profile of the gas mass (calculated from

In the following a detailed description of the design of the trajectories is given.

The central temperature was chosen randomly within the allowed temperature interval. Then the temperature of the next outer radius point was randomly chosen in a temperature "window" with a given size, which we will call $\delta$ in the following, and a specially defined centre position of the window. This centre position was defined by having the same ratio of distance to the lower temperature limit to the total size of the channel as the previous temperature at the last radius. The subsequent temperature points of the trajetory are determined in the same way. This "diffusive process" of constructing the temperature trajectories, guarantees a certain smoothness of these curves, which is characterized by the two parameters, stepsize ($\Delta r$), and window size ($\delta$). We chose the smallest possible stepsize, that provided solutions, $\Delta r=60$ kpc (the smaller the stepsize, the higher the gradients, the more difficult to get trajectories, which lead to physical results). $\delta=0.6$keV seemed to be the value that provided the best compromise between enough freedom for the trajectories in the selection of temperatures on the one hand and suppressing large oscillations on the other hand. Some typical trajectories are shown in Fig. 5.

Of course, the trajectories had to fulfill the condition of monotonically increasing total mass. To match this requirement, the temperature window for the selection of subsequent temperature points is restricted accordingly.

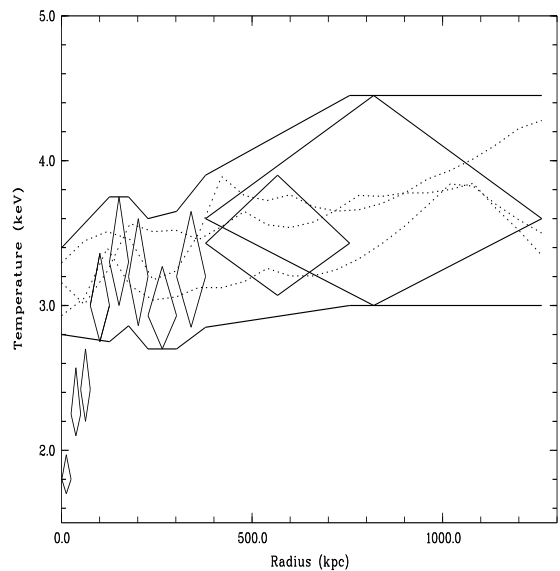

**Fig. 5.** Temperature channel used for the Monte-Carlo simulation. The diamonds indicate the errors in temperature on the $1\sigma$-level. Some possible and used trajectories (dashed lines) are also shown.

restriction, are discarded. A possible occurrence, for example, is a trajectory with a large negative gradient; then the mass was determined to be very high, according to the mass formula. To still provide the same mass, or even a higher one in the next step, the temperature had to decrease again by a large amount. If this decrease was no longer possible, for example if the temperature was already at the lower limit of the channel, the trajectory was ruled out and another one started. Fig. 6 shows the results for the constraints of the mass profiles for three different choices of the $\delta$-parameter. One notes that the results are very weakly dependent on this choice of $\delta$, which demonstrates the robustness of this method.

The results from this method were also compared to an approach analogues to that used by Hughes (1989) for the Coma cluster. This approach has the disadvantage of making a priori assumptions on the potential of the cluster taking the form of a modified isothermal sphere:

$$\rho_{grav}(r) = \rho_0 \left(1 + \frac{r^2}{a^2}\right)^{-\eta/2} \qquad (7)$$

where $\rho_0$, $a$ and $\eta$ are free parameters. For a representative set of these potentials we calculated the emission measure to the observations. The set of valid models covers the parameter range $a = 70...300$ kpc and $\eta = 1.75...3$. The upper and lower mass limits of these combined valid model set are also shown in Fig. 6.

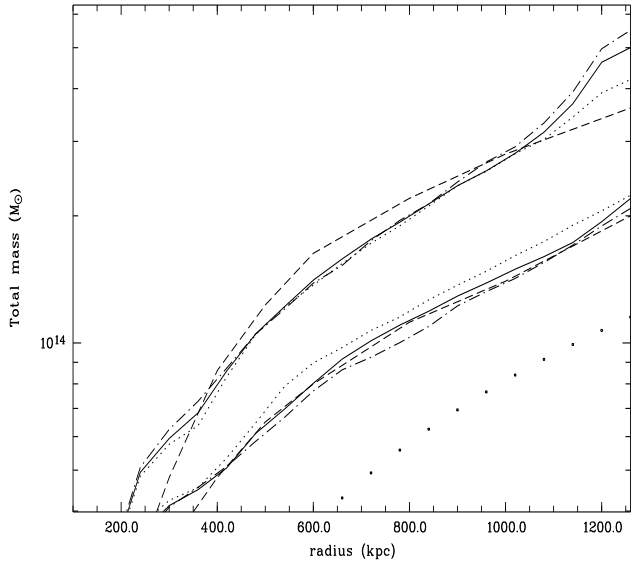

**Fig. 6.** Integrated mass profile of AWM7 derived with the Monte-Carlo method. The lines show the upper (90%) and lower (10%) limits of the simulation. The full line shows the mass profile with $\delta = 0.6$ keV, the dotted curves show the mass profile with $\delta = 0.2$ keV, and the dashed and dotted curves show the results for $\delta = 0.8$ keV. The dashed curves are the results of the method described by Hughes (1989). The dots indicate the gas mass.

If we extrapolate the Monte-Carlo method out to a radius of 1° (1.8 Mpc), we obtain a mass-to-light ratio of 200–550 $M_\odot/L_\odot$. This extrapolation is necessary to be able to use the luminosities derived by BGHLD. For calculating the mass profile up to this radius we had to constrain the simulation to a smaller $\delta$-value to get solutions ($\delta = 0.2$ keV with a stepwidth of 60 kpc). A smaller $\delta$ can lead to an underestimation of the errors spread by the temperatures but the $\delta$-value is still large enough that the trajectories fill the entire temperature "channel".

### 5.2. Cooling flow analysis

The mass profile determined from the X-ray data can also be used as input for the determination of the mass accretion rate in the cooling flow region. For cooling flow studies with pre-ROSAT X-ray data, the parameter for the central gravitational potential had to be guessed or to be taken from optical observations (e.g. Fabian et al. 1984; Arnaud 1988). The ROSAT data now allow the simultaneous calculation of the potential and the accretion rate from X-ray data.

For this analysis the $\beta$-model fit and the mass profile were determined for the inner region of the cluster. The accuracy of the $\beta$-model fit is demonstrated in Fig. 3. From the temperature and density distribution we derive a cooling flow radius of $\sim 60$-$80$ kpc, which is defined as the radius where the cooling time is $10^{10}$ y. For this analysis we use the temperature profile of the four innermost rings (from the one temperature fit). The mass accretion rate was determined from

$$\dot{M}(r) = 4\pi r^2 \frac{n_e n_H \Lambda(T) - \frac{5k\dot{M}_{outer}T}{8\pi r^2 \mu \Delta r m_p}}{\frac{5k}{\mu \Delta r m_p}(T_{outer} - 2T) + \frac{GM}{r^2}}. \qquad (8)$$

$\Lambda(T)$ is the volume emissivity per unit particle density, and a function of $T$. We calculated from the cooling flow radius inwards. "$Outer$" denotes the values of the next outer annulus and $\Delta r$ is the stepwidth in kpc. We determined this formula from energy conservation (the infall and cooling of matter must compensate the energy loss), and the fact that the condensed material provides energy for the emission process until it has reached low temperatures (for a detailed description of the cooling flow formulas see e.g. Arnaud 1988). In this calculation we neglected the effect of heat conduction. Eq.(8) was determined with a binwidth of 10 kpc. The resulting lower and upper limits for the mass accretion rates (again 10% and 90% levels as described for the total mass analysis) are shown in Fig. 7. Again, the projected temperature profile of Table 1 (first four annuli) were used. The mass accretion rate is decreasing with decreasing radius. At the cooling flow radius we ports the result of Edge et al. (1992) very well. For the error analysis we included the error of the mass profile, and as the errors of the density determination were also large, we included also the error of $\beta$, $a$, and the central density.

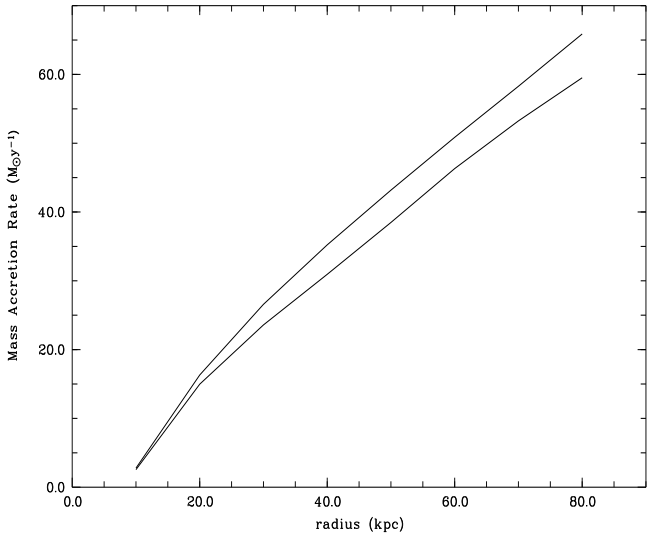

**Fig. 7.** The results for the cooling flow analysis with a step-width of 10 kpc. The calculated cooling flow radius is about 60–80kpc. We show the upper (90%) and lower (10%) limits.

## 6. Discussion

### 6.1. Morphology

The fact that AWM7 is elliptically elongated in the direction of the Perseus cluster and also follows the general orientation of the Perseus-Pisces chain of galaxies raises the question whether the elongation is caused by tidal interaction of material with the Perseus cluster or asymmetric accretion of material during the formation of the AWM7-cluster. Calculating the tidal effect of Perseus on the potential and the gas distribution yields a disturbance of the gas isodensity surface of about 2% at most. For this calculation we included a total mass of the Perseus cluster of $2 \times 10^{15} M_\odot$, a distance between both clusters of 8.4 Mpc, and a median mass of AWM7 of $3 \times 10^{15} M_\odot$. Thus the tidal distortion by the Perseus cluster cannot be the reason for the elongation of AWM7, since the fitted ellipses have a ratio between semi-major and semi-minor axis of about 80%. Therefore, we conclude that the ellipticity of AWM7 is the result of asymmetric infall of matter caused by its own gravitational potential, originating preferentially from directions along the Perseus-Pisces chain. If the very inhomogeneous, one can also easily explain a temporary centre offset of the cD galaxy, which should also not occur for tidal interaction. If the disturbance came from along the Perseus-Pisces filament, it would result in a displacement mainly perpendicular to the line of sight, having the cD galaxy in the centre of the redshift distribution but offset to the projected center, exactly as it is observed.

The disturbance must have had the unusual effect, however, to also turn the major axis of the cD almost perpendicular to that of the cluster, since in isolated elliptical clusters the elliptical elongation of the cD is usually aligned with the cluster shape (Allen et al. 1994).

It is not yet clear whether the offset of NGC1129 from the X-ray centre is due to a peculiar velocity of NGC1129 or simply a shift of the X-ray emission away from the centre of symmetry of the cluster. Malumuth (1992) suggests that NGC1129 is not at rest in the cluster potential, but other authors (e.g. BGHLD) see clear evidence that the central dominant galaxy is at rest (in the line-of-sight). The difference between the cluster mean velocity and the central dominant galaxy in AWM7 is 32 km s$^{-1}$ in the calculation of Malumuth (1992). However, this difference is not significant as it is within the error bars of the determined redshifts.

### 6.2. Mass-to-light ratio

BGHLD determined a very high mass-to-light ratio of over 850–1400 $M_\odot/L_\odot$, much higher than the result here. This could be due to the increasing velocity dispersion. The velocity dispersion is increasing with radius, as it was first seen by Hintzen (1980), and also later by BGHLD. This possibly indicates that the cluster is still disturbed by infalling galaxies, a feature which could lead to an overestimation of the total mass. BGHLD find a velocity dispersion along the line of sight of $\sigma_r = 684$ (+169, -97) km s$^{-1}$ within a radius of 0.5 Mpc (BGHLD use $H_0$=100 km s$^{-1}$ Mpc$^{-1}$). In the outer regions (0.5-1.8 Mpc), they find a $\sigma_r$ of 1015(+240,-140) km s$^{-1}$. The mass-to-light ratio in our calculation (200–550 $M_\odot/L_\odot$) is in good agreement with results for other clusters.

### 6.3. Gas to total mass ratio

The high ratio of gas to total mass (0.11-0.27) found for AWM7 is in line with the findings for a series of rich clusters (e.g. Briel et al. 1992; Böhringer 1993) and further strengthens the cosmological problem of the baryon excess in clusters (White et al. 1993). This problem consists of the fact that several requirements of the most popular cosmological model can hardly be fulfilled at the same time: (i) the matter composition in clusters reflect the matter composition of the universe, (ii) the mean density of the universe is close to the critical density, (iii) the limits of are for example $\Omega=0.05(\pm0.01)$ in the models by Walker et al. (1991). Thus the upper limit of 6% baryons in a critical density universe is difficult to reconcile with the high baryonic fraction in AWM7 since we do not expect that matter is discriminated while accreting onto clusters. The baryon excess that has mainly been found in the analyses of very rich clusters is here now confirmed for the case of a poor cluster of galaxies.

### 6.4. Cooling flow

A very interesting feature of the spectral analysis is the enhancement of cold hydrogen in the centre. The excess of condensed gas was already found in former analyses of cooling flow clusters observed by ROSAT (e.g. White et al., 1991; Allen et al. 1993). Also radio observations of AWM7 showed a large amount HI of $\simeq 10^9 M_\odot$ (McNamara et al. 1990). When we estimate the mass of this cold gas in the central cooling flow region we obtain about $10^9$ $M_\odot$. This is still not enough to explain all the accreted material as this should be about $10^{11} M_\odot$, if the cooling flow started after half a Hubble time, but it is a very strong indication for the existence of accreting material.

### 6.5. Central region

It is rather interesting to note the calculated cooling flow radius (60-82kpc), the strong temperature drop in the inner three temperature bins (r$\leq$76kpc), and also very roughly the area where we note an isophote shift to the west, all highlight essentially the same region in the cluster. These findings, the coincidence of the cooling flow radius with the temperature drop just confirm that the standard cooling flow model is approximately correct.

*Acknowledgements.* We like to thank the ROSAT team for providing the processed data and the EXSAS team for the tools for the data reduction and useful discussions. We also thank S. Schindler, C. A. Collins, and R. G. Bower for helpful discussions and useful corrections on the paper. HB. is supported by the Verbundforschung, project number 50 OR 93065.